\documentclass[fleqn,usenatbib]{mnras}

\usepackage{newtxtext,newtxmath}

\usepackage[T1]{fontenc}

\DeclareRobustCommand{\VAN}[3]{#2}
\let\VANthebibliography\thebibliography
\def\thebibliography{\DeclareRobustCommand{\VAN}[3]{##3}\VANthebibliography}

\usepackage{graphicx}	
\usepackage{amsmath}	
\usepackage{subcaption}

\newcommand{\Ptr}{P_\mathrm{trace}}

\newcommand{\Teff} {T_{\rm eff} }
\newcommand{\heii} {He {\sc ii}}

\newcommand{\glog} {\log \textsl{\textrm{g}}}

\title[Synthetic Spectroscopy for WD Classification]{Synthetic Spectroscopy for White Dwarf Classification: Addressing Label Uncertainty and Class Imbalance}

\author[Vincent et al.]{
Olivier Vincent, \thanks{E-mail: o.vincent@umontreal.ca}
Pierre Bergeron
and Patrick Dufour
\\
D\' epartement de Physique, Universit\' e\ de Montr\' eal, C.P. 6128, Succ.~Centre-Ville, Montr\' eal, Qu\' ebec, Canada\\
}

\date{Accepted XXX. Received YYY; in original form ZZZ}

\pubyear{\the\year{}}

\begin{document}
\label{firstpage}
\pagerange{\pageref{firstpage}--\pageref{lastpage}}
\maketitle

\begin{abstract} 
With the imminent data releases from next-generation spectroscopic surveys, hundreds of thousands of white dwarf spectra are expected to become available within the next few years, increasing the data volume by an order of magnitude. This surge in data has created a pressing need for automated tools to efficiently analyze and classify these spectra. Although machine learning algorithms have recently been applied to classify large spectroscopic datasets, they remain constrained by the limited availability of training data. The Sloan Digital Sky Survey (SDSS) serves as the current standard training set, as it provides the largest collection of labeled spectra; however, it faces challenges related to severe class imbalance and uncertain label consistency across different surveys.
In this work, we address these limitations by training histogram gradient-boosted classifiers on a synthetic SDSS dataset to identify six ubiquitous chemical signatures in the atmosphere of white dwarfs, and test them on 14,246 objects with SNR$>$10 SDSS spectroscopy. We show our approach not only surpasses human expert performance, but also enables subtype classification and effectively resolves label transferability issues. The methodology developed here is adaptable to any spectroscopic survey, providing a critical tool for the astronomical community.
\end{abstract}

\begin{keywords}
White Dwarfs -- Methods: Data Analysis -- Catalogues -- Surveys
\end{keywords}



\section{Introduction}
Approximately 97\% of all stars in our Galaxy will ultimately evolve into white dwarfs, the final remnants left after stars exhaust their nuclear fuel and shed their outer layers \citep[][and references therein]{Fontaine2001}. Large samples of white dwarfs are necessary to understand their properties, such as their spectral evolution as they age \citep[][and references therein]{Bedard2024}, the processes of energy transfer and modelling of their atmospheres \citep[e.g.,]{Koester2010, Bauer2018, Cunningham2019}.
These remnants are also invaluable for several astrophysical studies such as estimating the ages of stellar populations, investigating the composition of planetary material, and understanding the progenitors of Type Ia supernovae \citep{Althaus2010, Toonen2012, Veras2021}.

Around 10 billion white dwarfs populate the Milky Way galaxy \citep{Napiwotzki2009}, with 360,000 candidates recently identified using astrometric and photometric data from the \emph{Gaia} mission \citep{GF2021, Vincent2023}. Several spectroscopic surveys, including the Sloan Digital Sky Survey V \citep[SDSS-V;][]{SDSSV}, the Dark Energy Spectroscopic Instrument \citep[DESI;][]{DESI}, the WHT Enhanced Area Velocity Explorer \citep[WEAVE;][]{WEAVE}, and the 4-metre Multi-Object Spectroscopic Telescope \citep[4MOST;][]{4MOST}, are now following up on these candidates. Among these, SDSS-V stands out with its ambitious goal to provide medium-resolution spectroscopy for approximately 300,000 white dwarfs, a significant increase from the 40,000 currently catalogued objects. Furthermore, \emph{Gaia}’s low-resolution spectra with high signal-to-noise ratios (SNR) are expected to cover a majority of these candidates.

The need for automated white dwarf classification algorithms has become pressing with the recent publication of $\sim$100,000 low-resolution white dwarf spectra from \emph{Gaia} Data Release 3 \citep{GaiaDR3}. Supervised classification algorithms, trained on previous SDSS catalogues, have already been applied to this dataset, successfully classifying a subset with high confidence \citep{GarciaZamora2023,Vincent2024}. Unsupervised algorithms have also proven effective, particularly in identifying white dwarfs with significant metal pollution \citep{Kao2024}. However, two major challenges persist: the reliance on prior human classifications and the issue of class imbalance. The latter arises because around 80\% of known white dwarfs have pure hydrogen atmospheres, 7\% are pure helium, 8\% exhibit featureless spectra, while the remaining 5\% contains all other spectral types\footnote{Estimates are based on main spectral types counts from the Montreal White Dwarf Database \citep[MWDD;]{Dufour2017}.}. Objects with mixed spectral signatures are often too rare to support effective training for supervised methods or clustering in unsupervised analyses, limiting the classification of these multi-type spectra. Additionally, the dependency on SDSS-derived classifications, which constitute the largest pre-labeled spectroscopic datasets, can lead to inaccuracies when instrumental resolution or SNR varies significantly, introducing label noise that hinders classifier performance.

In this work, we address these two limitations by generating an extensive dataset of simulated spectra for training supervised classification algorithms. Our approach overcomes class imbalance by generating additional spectra for underrepresented classes and assigns new labels to account for variations in resolution and noise levels. We demonstrate that this method outperforms existing classification techniques, including manual classifications by experts. Additionally, spectra that diverge significantly from our simulations are flagged as potential outliers using unsupervised density estimators, allowing for further inspection. The flexibility of simulated datasets enables our approach to adapt to upcoming survey data.

This paper is organized as follows. Section \ref{sec:metho} outlines the simulation process, the chosen classification and outlier detection algorithms, and the training procedures. In Section \ref{sec:results}, we evaluate our approach on synthetic and real SDSS data. Section \ref{sec:disc} compares our results with previous classification methods, explores new potential detections, and discusses the limitations of our approach. Finally, concluding remarks are presented in Section \ref{sec:conc}.

\section{Methodology}\label{sec:metho}
\subsection{Spectroscopic Data Simulations}\label{sec:sims}
The training set was generated using a variety of model atmospheres, as outlined in Section 4.2 of \citet{Vincent2024}, to cover common types of white dwarf stars. In summary, we generated spectra for pure hydrogen \citep{Blouin2018, Tremblay2011, Bedard2020}, pure helium, hydrogen-helium mixtures \citep{Blouin2019, GBB2019, Bedard2020}, carbon \citep{Dufour2005, Coutu2019, Blouin2019_DQ}, and metal-polluted atmospheres \citep{Blouin2018_DZ, Coutu2019}. Additionally, hot DQ model atmospheres from \citet{Dufour2008} were used to simulate carbon-atmosphere spectra up to an effective temperature of $\Teff=24,000$~K. We uniformly sampled the parameter grids of each model and linearly interpolated between 20,000 and 70,000 high-resolution spectra, depending on the model parameter complexity. To simulate additional continuum-normalized, featureless white dwarf spectra (DC), we added noise to a constant line of value 1, as described in the following section. Table \ref{tab:grids} summarizes the parameter ranges and the number of spectra generated for each grid.

To replicate the resolution and noise of SDSS observations, we first convolved the pristine spectra with a Gaussian kernel with a full width at half maximum (FWHM) of 3~\r{A}, a well-tested approach to modelling SDSS spectra \citep{Bergeron2011, Tremblay2011}, and interpolated the flux to match the survey's wavelength sampling between 3842 and 7000~\r{A}. We assumed observations are dominated by photon noise in a high photon count regime, which is appropriate for most observations and can be modeled using additive Gaussian noise. We applied random Gaussian noise to each pixel\footnote{In this work, a pixel refers to a single wavelength-flux.} to achieve a median signal-to-noise ratio (S/N) ranging between 9 and 50 for each spectrum. This range was chosen to match the S/N of real SDSS spectra that are used to test this approach, described later in Section \ref{sec:sdss}. To further augment the dataset, we applied a random radial velocity shift drawn from a uniform distribution between $-$100 and 100 km~s$^{-1}$. Additionally, we introduced slight variations in continuum shape by selecting pixels in the 85th flux percentile within a running window of 100 pixels. We fit a fourth-order Chebyshev polynomial to these pixels, added noise between $-1$\% and 1\% to each polynomial coefficient, and multiplied the spectrum by the resulting curve.

Each simulated spectrum was assigned binary labels for the chemical species present, as determined by the model atmospheres. For instance, a pure hydrogen atmosphere would receive a positive label for hydrogen and a negative label for all other species, while a hydrogen-helium mixture would have positive labels for both elements. The complete list of labels, including hydrogen, neutral helium, ionized helium, carbon, calcium, and DC (no features), is presented in Table \ref{tab:grids}. We note that effective temperatures ($\Teff$) and carbon abundances within carbon-atmosphere grids cover a range that can include ionized ($\Teff \gtrsim 16,000$~K) and molecular ($\Teff \lesssim 10,000$~K) carbon. Due to the gradual transitions between these states, complete separation is not feasible; therefore, the carbon label primarily refers to neutral carbon, though instances of ionized and molecular carbon are also included in the training set.

Labeling spectra solely based on their model grid is insufficient, as random parameter sampling and noise addition can alter the visibility of spectral features. For example, a mixed hydrogen-helium atmosphere may exhibit lines for both species, only one, or none, depending on the abundance ratio, $\Teff$, and surface gravity. Instead of manual classification, we used Label Spreading, a semi-supervised algorithm implemented in \texttt{scikit-learn} \citep{sklearn}. We assigned labels to a core subset of spectra with confirmed spectral features, calculated a similarity matrix for all spectra, and assigned labels based on the highest similarity values. Spectra without a similarity score above 0.5 for any label were discarded. The parameter ranges for each core subset are presented in Table \ref{tab:grids}. We verified the labels by plotting the average spectrum for each label within each grid and visually inspecting a subset of spectra per label. The total number of spectra per label and the contributing model atmospheres are detailed in Table \ref{tab:grids}.

\begin{table}
    \centering
    \begin{tabular}{|p{1.3cm}|p{2.7cm}|p{1.15cm}|p{1.3cm}|}
         \hline
         \hline
         Grid & Parameter ranges & Number of spectra & Possible Labels\\
         \hline
         \hline
         Pure H & 
         $\Teff =$ [1500, 150,000] \\
         & $\glog =$ [6.5, 9.5] & 
         40000 & H, DC\\
         \hline
         Pure He & 
         $\Teff =$ [11,000, 150,000] \\
         & $\glog =$ [6.5, 9.5] & 
         40000 & He, \heii, DC\\
         \hline
         H+He & 
         $\Teff =$ [1500, 150,000] \\
         & $\glog =$ [6.5, 9.5] & 
         70000 & H, He, \heii, DC\\
         \hline
         C & 
         $\Teff =$ [5000, 24,000] \\
         & $\glog =$ [7, 9] \\
         & log C/He = [$-$7.5, 2] & 
         60000 & C, DC \\
         \hline
         Ca (He) & 
         $\Teff =$ [4000, 16,000] \\
         & $\glog =$ [7, 9] \\
         & log Ca/He = [$-$12, $-$7] & 
         20000 & Ca, He \\
         \hline
         Ca (H+He) & 
         $\Teff =$ [4000, 16,000] \\
         & $\glog =$ [7, 9] \\
         & log He/H = [$-$2, 5] \\
         & log Ca/He = [$-$12, $-$7] & 
         50000 & Ca, H, He \\
         \hline
         DC & - & 10,000 & DC \\
         \hline
         \hline
    \end{tabular}
    \caption{Overview of the synthetic spectra generated for the training set.}
    \label{tab:grids}
\end{table}

\subsection{Histogram-based Gradient Boosting Classification}
The Histogram-based gradient boosting classification algorithm \citep{Friedman2001} is an efficient and robust machine learning model suitable for classifying large datasets with moderate input sizes. The algorithm builds an ensemble of decision trees sequentially, where each tree is trained to correct the errors of the previous ones. This process minimizes the binary cross-entropy loss, defined as:

\[
L(y, \hat{y}) = - \sum_{i=1}^{n} \left( y_i \log(\hat{y}_i) + (1 - y_i) \log(1 - \hat{y}_i) \right),
\]

\noindent where \( y_i \) is the true label and \( \hat{y}_i \) is the predicted probability for the \(i\)-th example.

In the histogram-based approach, pixel values are grouped into a fixed number of bins, creating histograms from the spectra. In our case, we split the spectra into 255 bins with open intervals, except for the first and last bins, which are closed to include the minimum and maximum wavelengths. That is, each flux pixel is assigned to one of a fixed number of bins based on its value. During tree construction, the algorithm then aggregates gradients over these bins to efficiently determine split points. These histograms reduce computational cost and memory usage by grouping similar values, allowing the model to calculate gradients and determine optimal splits more efficiently. By iteratively adding trees based on these gradient calculations, the classifier converges toward an accurate model without requiring precise pixel-level splits, as would be necessary in a traditional gradient boosting classifier. This binning strategy significantly speeds up training and is highly scalable, making it suitable for high dimensionality of stellar spectra.

We used the \texttt{Scikit-learn} implementation with default parameters. We did not find any gains in performance by varying the learning rate, L2 regularization and maximum leaf node parameters. We trained one classifier per chemical label (see Table \ref{tab:classifiers}), in contrast to our our previous work \citep{Vincent2023} where a single classifier was trained for all selected classes. Training a single classifier per label offers greater control as to what features are used as input, and thus better interpretability.

\subsection{Data processing}\label{sec:train}
To train the classifiers, we randomly selected up to 20,000 simulated spectra for every unique combination of labels. The resulting number of positive labels for each chemical trace is shown in Table \ref{tab:classifiers}. These spectra are then shuffled and split into two sets, 80\% for the training and 20\% for the validation. 

\begin{figure*}
    \centering
    \includegraphics[width=\linewidth]{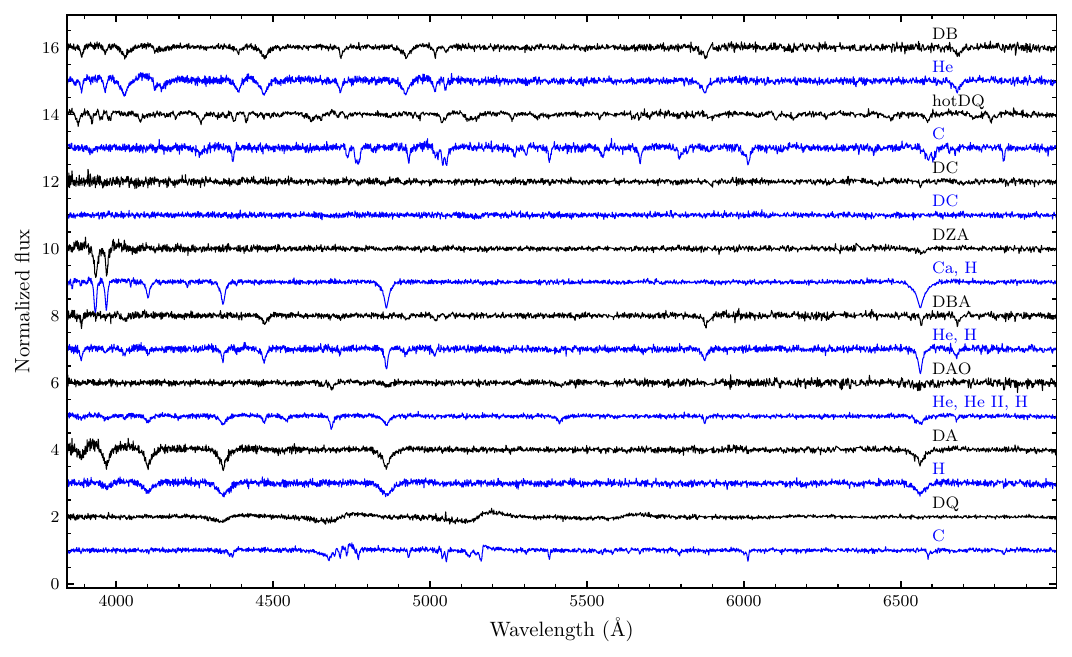}
    \caption{Examples of continuum-normalized spectra (blue) and real SDSS spectra (black) of similar classes. The spectroscopic classification or training label are shown above the real and simulated spectra, respectively.}
    \label{fig:simreal}
\end{figure*}

Prior to feeding the spectra to the classifiers, we normalized their continuum as follows. We first approximated the continuum using a moving median filter with a window size of 150 pixels ($\sim$135~\r{A}). We smoothed the curve obtained by the moving median with a 1D Gaussian filter with standard deviation of 10~\r{A}. We then normalized the spectra by dividing their flux with the smoothed curve. Examples of the continuum-normalized spectra are shown on Figure \ref{fig:simreal} along with real SDSS spectra for comparison.

The model atmosphere grids used to generate metal polluted spectra include a large variety of heavier elements. To ensure interpretability of the calcium classifier, we further trimmed the spectra to a maximum of 5000~\r{A} to emphasize the important to calcium lines around at 3934 and 3969~\r{A}. Similarly, spectra fed to the neutral carbon classifier were trimmed to 4500-5500~\r{A}, where the most prominent features reside. We also applied a dimensionality reduction step using the Scikit-learn implementation of Principal Component Analysis \citep[PCA;][]{Tipping1999} with 10 components. Swan bands in the atmosphere of white dwarfs tend to vary significantly in appearance \citep{Coutu2019, Blouin2019}, making the the optimal selection of discriminative pixels difficult. Dimensionality reduction alleviates this problem by making the classifier focus on global features.

\begin{table}
    \centering
    \begin{tabular}{|c|c|c|}
    \hline
    \hline
        Label & Number of spectra & Spectral region (\r{A})\\
    \hline
    \hline
        H & 132,034 & 3842-7000\\
    \hline
        He & 125,790 & 3842-7000\\
        \hline
        \heii & 43,379 & 3842-7000\\
        \hline
        C & 45,140 & 4500-5500\\
        \hline
        Ca & 40,130 & 3842-5000\\
        \hline
        DC & 48,455 & 3842-7000\\
        \hline
        \hline
    \end{tabular}
    \caption{Overview of labels and data for the training set.}
    \label{tab:classifiers}
\end{table}

\subsection{Detecting Outliers with Kernel Density Estimators}

Many machine learning algorithms tend to produce overconfident predictions when confronted with out-of-distribution data, often assigning very high or low prediction probabilities to inputs outside their training set. In this study, we assume that the spectra presented to the classifiers are those of white dwarf stars. This assumption is reasonable given that the selection criteria (e.g., color cuts on the \emph{Gaia} Hertzsprung–Russell diagram; \citealt{Vincent2023}, \citealt{GF2021}) effectively filter out non-white dwarf objects, resulting in a contamination rate of approximately 1\%. Thus, detecting spectra that fall outside the training set becomes a task of identifying white dwarfs that are not well-described by the model atmospheres used in our simulations. These outliers primarily include magnetic white dwarfs (MWDs), cataclysmic variables (CVs), binary systems with significant optical flux contributions from a main-sequence companion (WD+MS), and spectra with unusual noise effects.

To detect these outliers, we use a Kernel Density Estimator (KDE), a non-parametric method for estimating the probability density function of a dataset. By smoothing data points across a specified kernel, KDEs create a continuous estimate of the data’s underlying distribution. This KDE model, trained on our simulated spectra, allows us to calculate the likelihood that a real observation belongs to this estimated distribution, thereby identifying observations that deviate significantly. To decrease the number of dimensions the KDE has to learn, we first applied PCA to the training set spectra described in the previous section, reducing each spectrum to 150 principal components. We then trained the KDE using the Scikit-learn implementation, with a Gaussian kernel, a bandwidth estimated by the Scott method, and Euclidean distance. Both the PCA transformation and KDE parameters were saved for application to real observations.

\subsection{Sloan Digital Sky Survey Spectroscopy}
To evaluate our algorithms, we used spectra from the SDSS Data Release 18 \citep{SDSS18}, selecting those with a white dwarf probability $P_\mathrm{WD} > 0.6$ based on the machine learning classification from \citet{Vincent2023}. We used spectral classifications from the \citet{GF2021} \emph{Gaia}-SDSS catalog (referred to GF21 henceforth), which provide binary values for each possible label in Table \ref{tab:classifiers}. Spectra classified as type DO are assumed to exhibit both neutral and ionized helium features. For WD+MS binary systems, only the white dwarf classification is considered, as flux contributions from the main-sequence companion are generally not prominent at wavelengths shorter than 7000~\r{A}. We excluded spectra with spectral types such as GALAXY, MWD, N/A, STAR, UNKN, Unreli, WD, and WDpec from our test set due to the lack of comparable labels. Our selection totaled 14,246 objects with at least one spectrum at SNR$>$10 (using the SNR provided in GF21). The spectra were processed following the same procedure described earlier.

The spectral types in GF21 follow the traditional nomenclature proposed by \citet{Sion1983}, where white dwarf spectral types start with D (standing for degenerate), followed by one or more letter(s) indicating the presence of specific elements: A for hydrogen, B for neutral helium, Q for carbon, O for ionized (and neutral, if present) helium and Z for metals. Multiple letters indicate the presence of multiple elements in order of strength, for example, DBA indicates an atmosphere showing both neutral helium and hydrogen. Here, we assign a binary label for each of the 6 possible classes learned by the classifiers based on the spectral types in GF21. We assume D*O stars have both neutral and ionized helium and D*Z stars all show calcium lines.

\section{Results}\label{sec:results}
\subsection{Performance on Synthetic Data}\label{sec:synth}
\begin{figure}
    \centering
    \includegraphics[width=\linewidth]{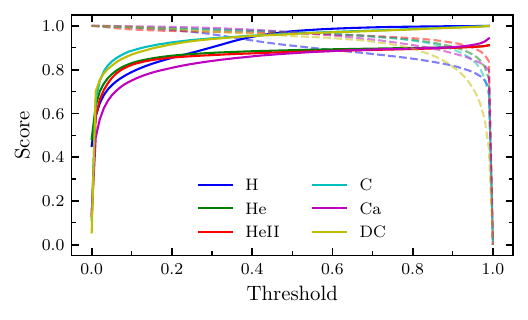}
    \caption{Precision (full lines) and recall (dashed lines) scores against classification probability threshold obtained from the synthetic validation set for each classification model.}
    \label{fig:PRcurves}
\end{figure}

To assess the effectiveness of our classification algorithms, we first validate their performance on a synthetic validation set in order to identify any potential biases or issues. We use the precision and recall metrics defined in \citet{Vincent2023} to quantify classification purity and the completeness, respectively, of the classifier predictions. Figure \ref{fig:PRcurves} presents precision and recall scores as a function of the classification probability threshold. Overall, the performance is very good, with precision scores reaching approximately 90\% for He, \heii, and Ca, and close to 100\% for H, C, and DC at a probability of trace detection ($\Ptr$) minimum threshold of $0.7$. Recall scores are around 95\% across all elements, except for H, which has a recall of 85\%.

Some misclassifications in the synthetic spectra result from label noise inherent to the data simulation process. Certain parameter combinations within specific model atmosphere grids can produce spectra lacking expected features. For instance, white dwarfs with mixed hydrogen and helium atmospheres, simulated at $\Teff > 40,000$ K with the model grids from \citet{Bedard2020}, are expected to show ionized helium features. However, extreme H/He abundance ratios may yield spectra resembling pure H or pure He atmospheres. We find that nearly all misclassified spectra originate from mixed-atmosphere model grids (e.g., grids with mixed H and He, or those including metals with H or He). Additionally, the added noise can obscure spectral features, making straightforward label assignment challenging.

To mitigate these issues, we applied the semi-supervised labeling approach described in Section \ref{sec:sims}, though a small proportion of erroneous labels persists. The impact of noisy labels on classifier performance has been studied extensively \citep{Natarajan2013, Frenay2014, Gosh2017}. Generally, noisy labels have minimal effect when they comprise less than 10\% of the dataset; however, noise levels around 40\% can significantly degrade classifier performance. We estimated the noise level in our training data by comparing classifier predictions for our validation set to the original model grids. If a detected chemical species could not be produced by the originating grid (e.g., calcium detected in a spectrum from a pure hydrogen grid), we categorize it as a noisy label. We estimate that approximately 7\% of our data contains noisy labels, primarily from grids with both hydrogen and helium atmospheres. This level of noise falls within the low-impact range. Additionally, larger training datasets have been shown to mitigate label noise effects \citep{Rolnick2017}, an advantage we leverage by generating additional synthetic data as needed.

Although classifier performance is high, it is useful to outline common sources of label noise and their potential impact on results. The most frequent labeling error arises from non-detection of hydrogen in mixed H and He atmospheres, accounting for 59\% of total misclassifications in the synthetic validation set. This issue primarily affects spectra generated with the H+He and Ca(H+He) grids, where hydrogen labels were either missing or incorrect. The second most common error involves helium misclassifications, constituting 21\% of total misclassifications. Similar to hydrogen errors, helium mislabels are found in edge-case spectra from the same grids. The remaining misclassifications are distributed across all labels without notable patterns.

\subsection{Spectroscopic Classification of SDSS Data}\label{sec:sdss}
We test the performance of our classification algorithms on the SDSS data described in Section \ref{sec:metho}, excluding spectra labeled as CV. A spectrum is classified as showing a given trace if its classification probability $\Ptr > 0.5$. In cases where both DC and another trace have high probabilities, we prioritize the other trace, ignoring the DC classification. We first evaluate classifier performance on 6924 spectra with SNR$>$20, where spectral features are typically clear. 

\begin{figure}
    \centering
    \begin{subfigure}{\linewidth}
        \centering
        \includegraphics[width=0.99\linewidth]{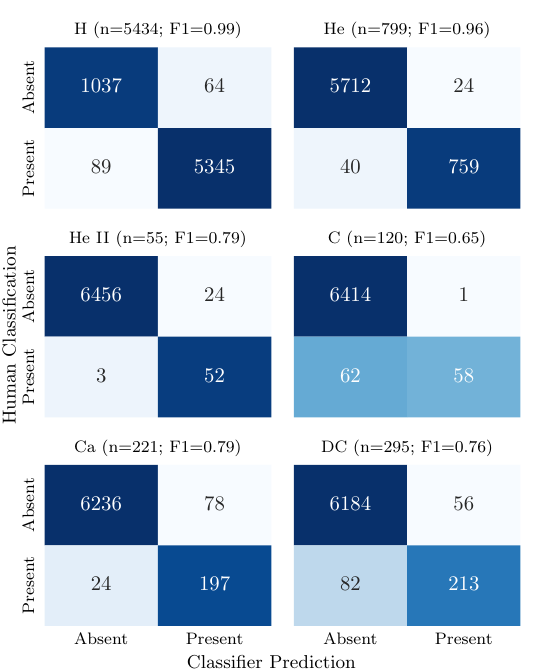}
        \text{(A)}
    \end{subfigure}
    \begin{subfigure}{\linewidth}
        \centering
        \includegraphics[width=0.99\linewidth]{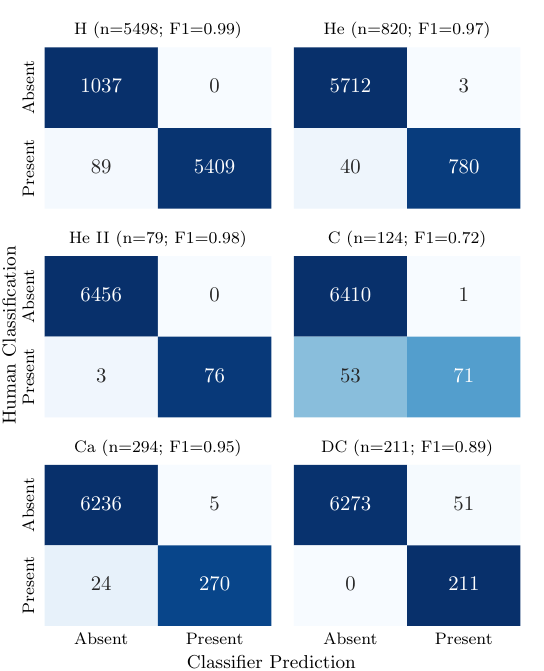}
        \text{(B)}
    \end{subfigure}
    \caption{Confusion matrices for each classifier applied on previously unseen SDSS spectra with SNR>20. The number of spectra and F1 score is shown above each matrix. Panel A shows the results for a classification threshold of $\Ptr>0.5$ across all classes, while panel B shows the results corrected for erroneous labels (see text). A $\Ptr>0.1$ threshold is used for the carbon classification in panel B.}
    \label{fig:cmatrix20}
\end{figure}

The confusion matrices for each classifier are presented on panel A of Figure \ref{fig:cmatrix20}. The y-axis shows human classifications, while the x-axis corresponds to classifier predictions. Each matrix cell shows the number of objects with their corresponding classifications. The number of true negatives (trace is absent according to both) is indicated in the top left cell, the number of (trace is present according to classifier, but absent according to human classification) in the top right cell, the number of false negatives (trace is present according to human classification, but absent according to classifier) in the bottom left cell and the number of true positives (trace is present according to both) in the bottom right cell. The cell colors reflect the percentage of objects within that cell for a given row, with darker shades indicating a larger percentage. Hence, a perfect agreement between human and classifier classifications would result in a diagonal matrix. We also display the F1 score, i.e. the harmonic mean of the precision and recall scores, above each matrix. The F1 scores represent the overall performance of the classifiers, assuming the human classifications are correct. We find F1 scores of 99\% for H, 96\% for He, 72\% for Ca, and 80\% for \heii, C, and DC.

Manual inspection of the discrepancies reveals the following trends. False positive detections for H, He, \heii, and C from the classifiers were found to be correct in each case, meaning the classifier correctly detected a trace that was not seen by human experts. Conversely, for false negatives, the human classification is always correct. Among 44 false negative H detections, 34 originate from mixed-atmosphere white dwarfs (DAB/OZ), 4 from magnetic DA, 2 from WD+MS systems with substantial companion contamination, and 4 from DA. For false negative helium detections, 13 are from mixed-atmosphere (DBA/OZ), 5 from DBH, 2 from DB, 1 from a DB+DC system, and 3 from DAO cases assumed to have neutral helium due to the presence of ionized helium, although neutral helium is absent. These findings align with the performance tests on synthetic data, indicating small performance issues when detecting hydrogen or helium in mixed atmospheres. In many cases, the false negatives are edge cases where spectral features are near the limit of detection.

A similar trend is observed for DC classifications, where classifiers accurately identify non-DC spectra that are erroneously labeled as DC in the GF21 catalog. In cases where classifiers produce false positive DC classifications, these spectra are also found not to be true DCs. Of the 47 false positives, 23 are cool DQ spectra with subtle features, 6 are magnetic white dwarfs, 5 are WD+MS systems with significant contamination, and the remaining are noisy DA and DB spectra. For carbon classifications, we find 54 false negatives, including 15 hot DQs (not included in the training set) and 39 cool DQs. By lowering the probability threshold to $\Ptr > 0.1$ for carbon, we correctly classify 19 additional cool DQs previously missed, with only two false positives (a WD+MS and a DAH).

Calcium classifications present additional complexity. Among 178 false positives, 150 spectra turned out to have calcium features in opposition to the human classification, while 28 do not show clear Ca {\sc ii} $\lambda$3969, Ca {\sc ii} $\lambda$3934, or Fe {\sc ii} $\lambda$4233 lines. Of these, 16 are DBA with faint hydrogen features, and 11 are classified as DB in GF21. As indicated by the synthetic validation results (see Section \ref{sec:synth}), spectra with hydrogen, helium, and calcium tend to be misclassified as mixed hydrogen-helium atmospheres during the semi-supervised labeling process. We suspect that many synthetic spectra generated from metal-polluted grids were labeled as containing calcium, even when they visually resemble pure helium or mixed H and He atmospheres. This overlap likely contributes to classifier challenges in distinguishing calcium from helium features, particularly in regions where these features coincide (e.g., $\lambda3969$). The remaining false positives include 1 hot DQ, 5 DB, 2 DA, and 4 DCs with increasing noise at blue wavelengths.

A total of 78 spectra do not exceed any classification probability threshold $\Ptr > 0.5$. These mostly include spectra outside the training distribution, such as 45 WD+MS, 3 hot DQ, 2 DAH, and 1 DAQ. The remaining spectra are highly noisy, comprising 18 DC, 6 DA, 2 DB, and 1 DQ.

In summary, classifier detections are nearly always accurate at a threshold of $\Ptr > 0.5$ for spectra with SNR$>$20, with only a few missed edge cases. The classifier precision is nearly 100\% for all traces except DC, which has a precision of 80\%. This threshold effectively yields highly pure samples of white dwarfs, albeit with a minor completeness trade-off. Panel B of Figure \ref{fig:cmatrix20} shows updated confusion matrices with corrected true labels based on our manual inspection. The corrected F1 scores are above 90\% for all traces (including carbon if hot DQs are excluded), consistent with the synthetic validation performance.

\begin{figure}
    \centering
    \begin{subfigure}{\linewidth}
        \centering
        \includegraphics[width=\linewidth]{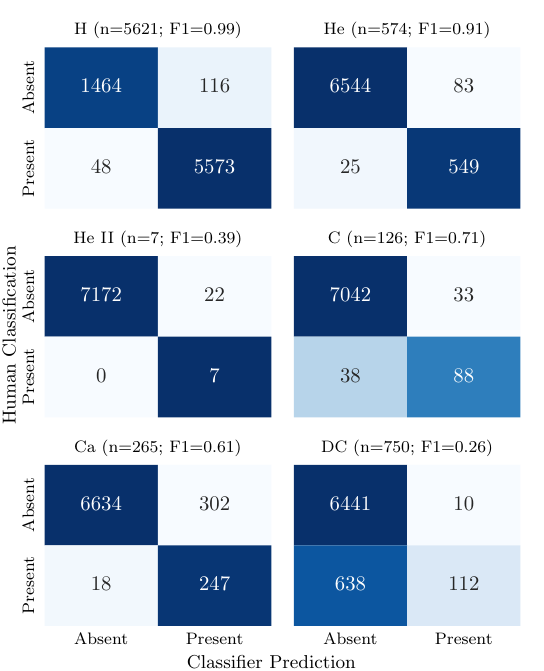}
        \text{(A)}
    \end{subfigure}
    \begin{subfigure}{\linewidth}
        \centering
        \includegraphics[width=0.99\linewidth]{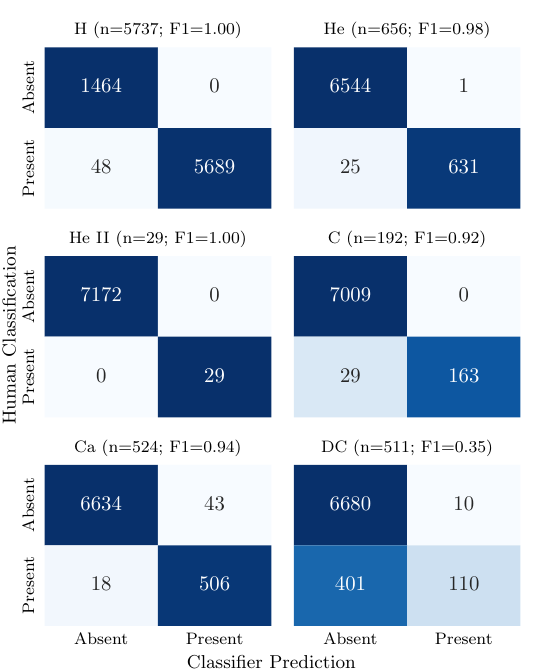}
        \text{(B)}
    \end{subfigure}
    \caption{Confusion matrices for each classifier applied on previously unseen SDSS spectra with 10>SNR$\leq$20. The number of spectra and F1 score is shown above each matrix. Panel A shows the results for a classification threshold of $\Ptr>0.8$ across all classes, while panel B shows the results for corrected labels (see text). A $\Ptr>0.4$ threshold is used for the carbon classification in panel B.}
    \label{fig:cmatrix10}
\end{figure}

We now analyze the 7343 spectra with 10$<$SNR$\leq$20. At lower SNR, SDSS observations more frequently deviate from the noise model used in our simulations, particularly from the assumption that noise is wavelength-independent. To reduce false positives arising from this variation, we set a higher threshold of $\Ptr > 0.8$. Figure \ref{fig:cmatrix10} presents the resulting confusion matrices. The F1 scores for hydrogen, helium, and carbon are similar to those at higher SNR before label corrections. However, the scores for ionized helium, calcium, and DC are significantly lower, at 39\%, 61\%, and 26\%, respectively. These reduced scores largely stem from an increase in false positives, which, as noted at higher SNR, tend to be true detections. A total of 486 spectra fail to exceed the classification threshold, 421 of which are labeled as DC in the GF21 catalog.

To verify whether discrepancies follow similar patterns as in the SNR$>$20 regime, we randomly select up to 100 false positive and 100 false negative classifications per trace and inspect them manually. The findings remain consistent across all traces except DC. 401 spectra classified as DC in the GF21 catalog do not exceed any probability threshold, accounting for the majority of the 638 false positives. This suggests that many current DC classifications deviate from the idealized model of a straight line with added noise, possibly due to low-level signals within the noise or an inaccurate noise model for low-SNR observations. A preliminary exploration of this issue is conducted in the following section.

To estimate the upper performance limits of the classifiers in the 10$<$SNR$\leq$20 range, we apply corrections to the labels based on insights from the higher SNR results. We assume false positives are correct for hydrogen, helium (neutral and ionized), and carbon. All false negative DC classifications are set as correct, except for the 401 spectra without any classification. Calcium false positives are marked as correct, except for spectra with a DBA/DAB spectral type in GF21 or if both calcium and helium are detected, as this combination often corresponds to DBA with weak hydrogen features. The threshold for carbon detection is adjusted to $\Ptr > 0.4$. The corrected confusion matrices are shown in Figure \ref{fig:cmatrix10}. Performance improves substantially, with F1 scores above 90\% for all traces except DC at 35\%. If non-classified objects are counted as DCs, as in GF21, the DC classifier performance also exceeds 90\%.

\begin{figure}
    \centering
    \includegraphics[width=\linewidth]{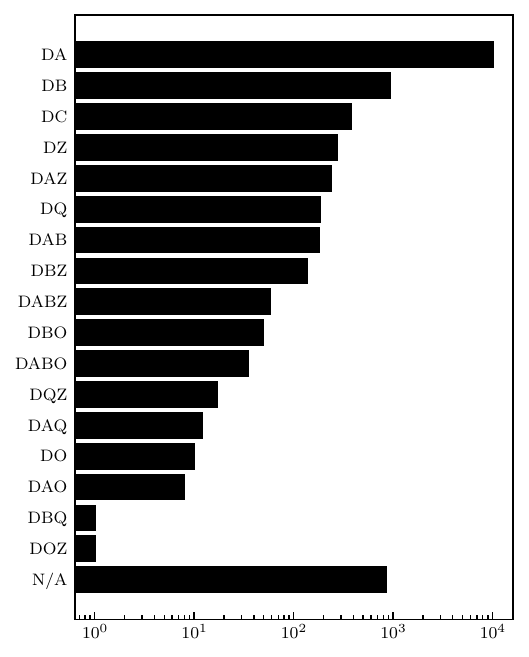}
    \caption{Number of spectra per spectral type obtained with our classifiers. Note that these designations are sorted alphabetically and do not reflect feature strength, and the O tag refers to ionized helium only (see Section \ref{sec:sdss}).}
    \label{fig:counts}
\end{figure}

The final classifications for all 13,941 objects are shown in Figure \ref{fig:counts}, displaying the count for each unique label combination. We rename the labels following the traditional nomenclature proposed by \citet{Sion1983} with two exceptions: 1) our O label indicating ionized helium only (hence, a DO would not show neutral helium), and 2) labels are sorted alphabetically, therefore do not reflect feature strength. We note that this convention may lead to minor inconsistencies with some literature, such as PG1159 being designated as DAO here. Spectra without classification probabilities above the threshold for their SNR range are labeled as N/A.

\subsection{Outlier detection within SDSS data}\label{sec:outliers}
We now apply the KDE outlier detection method, described in Section \ref{sec:metho}, to our SDSS sample. The goal of this method is to flag data that is too different from the training distribution. Outliers are defined as white dwarf atmosphere types not included in the synthetic training set or with noise significantly different from the simulations. Spectra with unusual noise cannot be easily labelled as outliers and likely represent the majority of cases, and while some classes (e.g., hot DQ) were not explicitly included in the training set, they share many spectral features with other classes that were included (e.g., warm DQ) and are not appropriate to test outlier detection. We use spectra with magnetic (D*H) and CV spectral types from GF21 as proxies to evaluate outlier detection performance, as they are the only two types guaranteed to be outside the training set. 

\begin{figure}
    \centering
    \includegraphics[width=\linewidth]{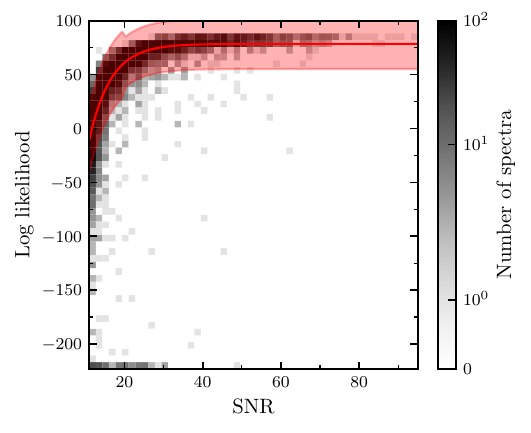}
    \caption{SDSS spectra counts as a function of signal-to-noise ratio and log-likelihood of belonging to the distribution learned by synthetic spectra KDE. Spectra are tagged as outliers if they fall outside the red zone representing one (SNR>20) or two (SNR$\leq$20) standard deviations around the fitted exponential curve (red line).}
    \label{fig:LLH}
\end{figure}

To compute the probability of SDSS spectra belonging to the flux distribution of synthetic spectra, we use the PCA model trained on synthetic data to reduce their number of dimensions and compute the log-likelihood of the distribution learned by the KDE. We show the log-likelihood as a function of SNR in Figure \ref{fig:LLH}. For visibility, any log-likelihood value smaller than $-225$ is set to $-225$. We find a clear correlation between the SNR and the likelihood of SDSS spectra, closely following a negative exponential trend. The likelihood of all SDSS spectra significantly drops at low SNR, increasing until SNR=30 and plateauing there. Many spectra appear to have much smaller likelihoods than the average for a given SNR bin, indicating a high probability of being an outlier. We also find a large accumulation of SNR<30 spectra with a log-likelihood lower than $-225$. 

The correlation between SNR and likelihood supports the idea that the noise model used in our simulations departs from the noise affecting SDSS spectra at low SNR. Classification performance does not appear to be significantly affected, as shown by the performance metrics in the previous section. We reiterate that classifiers do not know the difference between real spectral features and noise that significantly differs from the training set, and so false positives are expected to drastically increase with low SNR values, which is also what we found by varying the classification threshold. 

We identify outliers by selecting spectra that have significantly different likelihood values than the rest for a given SNR regime. To do so, we fit an exponential function to the log-likelihood against SNR, compute the residuals between the curve and each data point, and compute the standard deviation of the residuals for the two SNR regimes. The spectra with SNR>20, with select all spectra with residuals greater than one standard deviation as outliers, while for lower SNR, we select those with residuals greater than two standard deviations. The fitted curve and selection regions are shown on Figure \ref{fig:LLH}. 294 out of 13,763 spectra are tagged as outliers, with half of them from the outlier spectral types defined above. More specifically, 92 out of 100 CV and 49 out of 136 D*H were correctly selected, resulting in a recall of 60\%. Manual inspection of the other 153 reveals that they are indeed outliers, containing a wide variety of unusual features such as heavy pollution, contamination from a main sequence companion, large data reduction artifacts and large increases of noise in the red- or blue-most parts of the spectra. A gallery of randomly selected outliers is shown in Appendix \ref{app:outliers}. Overall, the outlier detection method applied here provides a qualitatively satisfactory performance.

\section{Discussion}\label{sec:disc}
\subsection{Comparison with Human Performance}
In this work, we created a completely new database of synthetic spectra to train automated classifiers, eliminating the need for human pre-labeled datasets. This approach makes our classifiers fully independent of previous classification results, contrasting with the classifiers in \citet{Vincent2023}, which rely on existing SDSS classifications and, to some extent, are trained to replicate them. In this section, we compare the performance of our classifiers to that of human classification (GF21) and neural networks trained on SDSS data \citep{Vincent2023} by evaluating discrepancies in classifications.

\begin{figure}
    \centering
    \includegraphics[width=\linewidth]{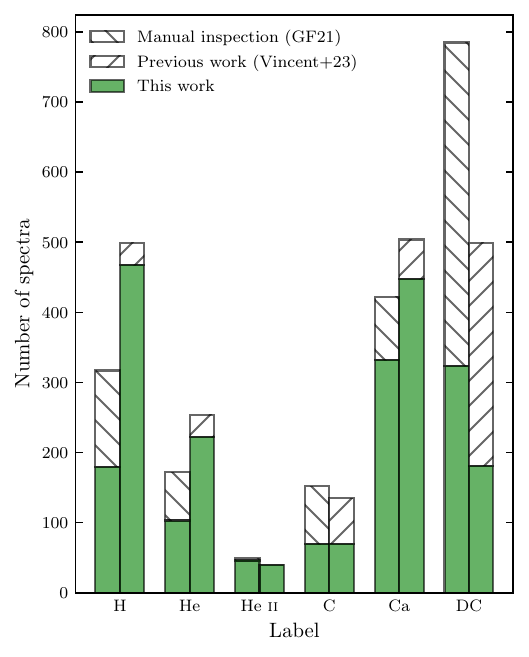}
    \caption{Number of discrepancies between our classifications and the results of GF21 and \citet{Vincent2023}. Green bars represent the number of instances where classifications in this work are correct, while hatched bars indicate previous classifications were correct.}
    \label{fig:correct}
\end{figure}

Discrepancies between our classifications and those obtained through visual inspection were discussed in Section \ref{sec:sdss}. Figure \ref{fig:correct} presents a histogram showing the number of discrepancies for each label, including all spectra, alongside the number of correctly classified spectra by either our classifiers or previous manual inspection. Our classifiers outperform manual inspection for most labels, correctly classifying 57\% of hydrogen discrepancies, 61\% for neutral helium, 94\% for ionized helium, and 80\% for calcium. For carbon and DC labels, performance is similar to manual inspection, with correct classifications in 48\% and 41\% of discrepancies, respectively. Notably, most DC discrepancies occur in low-SNR spectra with noise characteristics that deviate from our simulation assumptions, resulting in increased classifier uncertainty as SNR decreases (see Figure \ref{fig:LLH}). These spectra did not meet the classification probability threshold for any label, including DC, and therefore do not impact the precision of other labels. Additionally, the carbon label encompasses hot DQ and peculiar DQ classifications from GF21, two atmosphere types not explicitly included in our training set.

To compare our classifications with those from \citet{Vincent2023}, we assume that the true labels are those obtained in this work if they agree with classifications from GF21, along with the visually verified discrepancies discussed in Section \ref{sec:sdss}. For all other spectra, we defer to the classifications in \citet{Vincent2023}. Discrepancies are displayed in Figure \ref{fig:correct}. Our classifier is correct in 94\% of hydrogen classifications, 87\% for neutral helium, 100\% for ionized helium, and 89\% for calcium. Similar to the human performance comparison, the accuracy drops to 52\% for carbon and 36\% for DC. Although this comparison is somewhat limited because the neural network classifiers in \citet{Vincent2023} were designed to identify only the main atmospheric chemical trace, it still offers a useful benchmark for models trained on real data. The difference is particularly pronounced for the DC label, where classifiers trained on real observations can confidently identify featureless spectra, even with varying levels of noise. It is worth noting that our new approach correctly identified all traces in 674 out of 767 spectra with two or more traces, a significant upgrade from previous classifiers that were restricted to solely detecting the dominant trace.

These results demonstrate that synthetic training sets can be used to develop classifiers that outperform both human inspection and neural networks trained on SDSS data. This approach addresses the challenge of highly imbalanced white dwarf classes. For instance, our training set contains 43,379 spectra with ionized helium (see Table \ref{tab:classifiers}), whereas fewer than 200 such spectra are currently known \citep{Bedard2020, GF2021} among the approximately 40,000 white dwarfs in SDSS. Our classifier outperforms previous methods by 80–100\% for this category. Despite the limitations of imperfect noise models and labeling, the extensive number of synthetic spectra provides a highly effective solution for white dwarf spectroscopic classification.

\subsection{New Detections}
In this section, we examine notable detections from Figure \ref{fig:counts}, excluding spectra identified as outliers. We compare these classifications to the information available in the MWDD to identify potential new detections.

Our classifiers revealed a substantial number of calcium detection discrepancies compared to previous studies, with a significant fraction being verified as true detections (see Figure \ref{fig:correct}). We selected 291 objects for which our classifier reports a calcium detection, whereas both GF21 and \citet{Vincent2023} do not. Objects identified as false positives for calcium detection during our visual inspection, as discussed in Section \ref{sec:sdss}, are excluded from this selection. According to classifications in the MWDD, 22 of these objects are already known to exhibit metal pollution, leaving 270 as potential new calcium detections. We applied the same selection process to hydrogen, helium, and carbon detections, identifying 55, 18, and 33 possible new cases, respectively. No new ionized helium detections were found.

In addition to single-element detections, we identified 15 DQZ white dwarfs that are currently classified as DQ, DZ, or DC in the MWDD. We also find two DBQ, \emph{Gaia} DR3 2732459327587247360 and \emph{Gaia} DR3 100567105912890496. The former turned to be a magnetic white dwarf according to classifications on the MWDD, while the latter is indeed a white dwarf showing both carbon and features at typical helium line positions. However, the helium features for 100567105912890496 are likely to be noise, as its effective temperature is too low to produce genuine helium features \citep[$\Teff=	5867\pm54$~K,][]{Caron2023}

Given that the SDSS white dwarf sample has undergone extensive characterization over the years, the likelihood of true new detections is diminishing, though still possible. The detections reported here are expected to be objects near the detection limit, unusual cases, or outliers that were not filtered out, each warranting careful inspection.

\subsection{Limitations}
While simulating large datasets for training classification algorithms achieves state-of-the-art performance, this approach also has notable limitations. We discuss the primary challenges below.

1. Incomplete Representation of White Dwarf Atmospheres. Simulating the full diversity of white dwarf atmospheres is challenging due to the wide variety of types. Notable examples include magnetic white dwarfs, which are difficult to model due to the numerous parameters describing the strength and geometry of their magnetic fields \citep{Hardy2023a, Hardy2023b}, WD+MS binary systems, which require modeling an extensive range of MS companion types \citep{Echeverry2022}, and CV systems, for which spectral features are not yet well-predicted by theoretical models \citep{Zorotovic2020}. In this study, we identify these objects as outliers without attempting classification. Dimensionality reduction techniques, such as UMAP \citep{UMAP}, offer a promising alternative for identifying these objects without requiring labeled training sets. UMAP has already been successfully applied to \emph{Gaia} XP spectroscopy to identify diverse white dwarfs with extreme metal pollution \citep{Kao2024}. However, as shown by Kao et al., unsupervised learning may struggle to form distinguish clusters of rare spectral types (e.g., DO and DB are mixed together at the tail of the DA cluster) if they are not numerous enough.

2. Complexity of Accurate Noise Modeling. Accurately modeling noise is challenging. In this study, we simulated spectra under the assumption that Gaussian readout noise dominates. However, we observed that SDSS spectra deviate from this model as SNR decreases (see Figure \ref{fig:LLH}), necessitating a higher classification threshold to reduce false positives. Ideally, a pipeline that accurately replicates instrumental and observational effects would improve interpretability and control, but it is highly time-intensive. An emerging alternative is foundation models \citep{Bommasani2021}, which can be pre-trained on spectroscopic data from multiple surveys and fine-tuned for specific tasks such as white dwarf classification. This approach implicitly learns noise models and has proven effective in areas like extragalactic astronomy \citep{Parker2024}, physical science modeling \citep{Nguyen2023, McCabe2023}, and recently, stellar spectroscopic data \citep{Leung2024}. Another potential solution is to use dimensionality reduction techniques to help classifiers focus on key spectral features, as we did with PCA for the carbon classifier (see Section \ref{sec:train}). However, these methods are also sensitive to noise, and further investigation would be needed to confirm performance improvements, which is beyond the scope of this study.

3. Difficulty in Labeling Edge Cases in the Training Set. As discussed in Section \ref{sec:synth}, our label propagation approach for synthetic spectra led to missing hydrogen labels in mixed hydrogen-helium atmospheres, as well as metals, resulting in false negative hydrogen detections and false positive calcium detections in real SDSS data. A potential solution might involve expanding the parameter limits of model atmosphere grids and improving the sampling of ambiguous parameter combinations. Nevertheless, some label noise is inevitable. Fuzzy labeling \citep{Kuncheva2000}, where label values range between 0 and 1 based on confidence rather than binary labels, has shown robustness against label noise \citep{Keller1985, Thiel2008} and may be worth exploring in future work.

4. Lack of Feature Strength Ordering in Labels. A minor limitation of our approach is its inability to sort labels by feature strength, as is commonly done in white dwarf classification \citep{Sion1983}. Determining what constitutes "visual strength" for a chemical species—whether it is the number of lines, their depth, or their equivalent width—remains a subject of debate within the community. We do not attempt to establish our own criteria here; rather, we advocate for a shift toward a simplified classification system where species are either detected or not, moving away from subjective interpretations.

\subsection{Catalogue}
The results presented in Section \ref{sec:sdss} are made available as a catalogue. Objects that were removed from our analysis due to uncertain spectral types in GF21 are also classified and included. The columns of the catalogue are listed in Table \ref{tab:cat}. We include the spectral types using the nomenclature defined at the end of Section \ref{sec:sdss}. Spectra with an outlier warning are indicated with an asterisk as the first character of the spectral type (e.g., *DA). The log-likelihood of belonging to the synthetic data distribution from Section \ref{sec:outliers} is also included.

Fine-grained selection of spectra can also be achieved using the classification probabilities listed as \texttt{prob\_X} in the catalogue, where \texttt{X} refers to the labels listed in Table \ref{tab:classifiers}. Thresholds can be decreased to increase the completeness of the selection at the cost of sample purity. Based on the verifications done in this paper, we advise the following: 

\begin{itemize}
    \item Particular caution when decreasing the selection threshold for spectra with SNR<20, as the number of false positive increases rapidly with lower threshold values;
    \item Very low threshold probability ($\Ptr<0.4$) for a complete sample of carbon detections;
    \item For white dwarfs with both hydrogen and helium (along with other species, if applicable) in their atmosphere, complementing our results with other catalogues will provide more complete samples;
    \item Spectra with SNR<20 and an outlier flag should be discarded for pure samples, as their noise might cause false positive detections.
\end{itemize}

We also include columns \texttt{pred\_X} containing a binary value indicating the presence of each species based on the detection thresholds discussed in Section \ref{sec:sdss}.

\begin{table*}
\caption{Columns in the catalogue of classifications. The full table is available as supplementary material.}
\label{tab:cat}
\centering
\begin{tabular}{ll}
\hline
\hline
Column Header & Description \\ 
\hline
\hline
    \texttt{{source\_id}} & \emph{Gaia} DR3 source identification number\\ 
    \texttt{{plate}} & SDSS spectrum plate number\\ 
    \texttt{{mjd}} & SDSS spectrum observation date\\ 
    \texttt{{fiber}} & SDSS spectrum fiber\\ 
    \texttt{{SNR}} & SDSS spectrum signal-to-noise ratio\\ 
    \texttt{{phot\_g\_mean\_mag}} & \emph{Gaia} photometric G magnitude\\ 
    \texttt{{phot\_bp\_mean\_mag}} & \emph{Gaia} photometric BP magnitude\\ 
    \texttt{{phot\_rp\_mean\_mag}} & \emph{Gaia} photometric G magnitude\\ 
    \texttt{{parallax}} & \emph{Gaia} parallax measurement\\ 
    \texttt{{spectype}} & Spectral type from this work (see Section \ref{sec:sdss})\\ 
    \texttt{{spectype\_VBD2024}} & Spectral type from \citet{Vincent2024}\\ 
    \texttt{{spectype\_GF21}} & Spectral type from \citet{GF2021}\\ 
    \texttt{{prob\_H}} & Probability of hydrogen detection\\
    \texttt{{prob\_He}} & Probability of neutral helium detection\\
    \texttt{{prob\_HeII}} & Probability of \heii detection\\
    \texttt{{prob\_C}} & Probability of carbon detection\\
    \texttt{{prob\_Ca}} & Probability of hydrogen detection\\
    \texttt{{prob\_DC}} & Probability of continuum-only spectrum\\
    \texttt{{loglikelihood}} & Log-likelihood of spectrum belonging to synthetic training set (see Section \ref{sec:outliers})\\ 
    \texttt{{outlier}} & Flag indicating whether spectrum is significantly different from training data (see Section \ref{sec:outliers})\\ 
\hline
\hline
\end{tabular}
\end{table*}

\section{Conclusions}\label{sec:conc}
In this study, we demonstrated the effectiveness of generating extensive synthetic datasets for training classification algorithms on white dwarf spectra. Our approach achieves high levels of purity (>95\%) and completeness (>90\%, with the exception of DC types) in detecting a range of spectral features, including hydrogen, neutral helium, ionized helium, carbon, calcium, and featureless spectra.

This work addresses two major limitations of existing classification techniques. First, it mitigates class imbalance issues by allowing for the targeted generation of spectra for underrepresented classes. It can identify every trace of white dwarfs with multiple spectral signatures with 88\% completeness, a task that previously very challenging to accomplish due to the small number of available SDSS labels. In principle, this could be extended to certain types that are currently lacking from the training set, such as magnetic white dwarfs. More importantly, the methodology presented in this work operates independently of prior human classifications, thus avoiding the propagation of potential errors due to different survey resolutions, noise levels, and spectral variability.

Our classification methodology has been shown to outperform expert visual inspections, offering a more reliable and scalable approach for large datasets. However, certain limitations remain, particularly in the accuracy of noise modeling at low SNR levels and completeness of carbon trace detection, which will be addressed in future work.

To detect outlier spectra, we employed kernel density estimation to calculate their likelihood of belonging to the synthetic training set. This method successfully identifies both spectral classes absent from the training data, such as cataclysmic variables and magnetic white dwarfs, and spectra with noise characteristics that diverge from our data simulation assumptions. Furthermore, unusual but interesting spectra can be flagged for further examination based on their combination of high signal-to-noise ratios (SNR) and low likelihood scores.

Our approach is particularly timely, aligning well with the eminent data releases from SDSS-V, DESI, \emph{Gaia}, and 4MOST, and offering a robust framework for advancing white dwarf classification in future large-scale surveys.

\section*{Acknowledgements}
We are grateful to the anonymous referee for a careful reading of our manuscript and for several constructive comments that helped to improve this paper significantly. 
The authors wish to thank Antoine B\'edard for providing models for hot white dwarf atmospheres, as well as Simon Blouin and Antoine B\'edard for useful discussions that improved the quality of this manuscript. This work is supported in part by NSERC Canada, the Fund FRQ-NT (Qu\'ebec), and the Centre de recherche en astrophysique du Qu\'ebec (CRAQ).

This work presents results from the European Space Agency (ESA) space mission \emph{Gaia} and Sloan Digital Sky Survey. \emph{Gaia} data are being processed by the \emph{Gaia} Data Processing and Analysis Consortium (DPAC). Funding for the DPAC is provided by national institutions, in particular, the institutions participating in the \emph{Gaia} MultiLateral Agreement (MLA). Funding for the Sloan Digital Sky Survey (\url{https://www.sdss.org}) has been provided by the Alfred P. Sloan Foundation, the U.S. Department of Energy Office of Science, and the Participating Institutions. SDSS-IV acknowledges support and resources from the Center for High-Performance Computing at the University of Utah, and is managed by the Astrophysical Research Consortium for the Participating Institutions of the SDSS Collaboration. This research has also made use of the NASA Astrophysics Data System Bibliographic Services.

\section*{Data Availability}
The catalogue presented in Table \ref{tab:cat} is made available as online material on the publisher website. It will also be made available as a downloadable file\footnote{www.astro.umontreal.ca/{\textasciitilde}ovincent/catalogues} and on the Montreal White Dwarf Database\footnote{montrealwhitedwarfdatabase.org} upon publication.



\bibliographystyle{mnras}
\bibliography{bib} 


\appendix

\section{Outlier spectra gallery}\label{app:outliers}
\begin{figure*}
    \centering
    \includegraphics[width=0.97\linewidth]{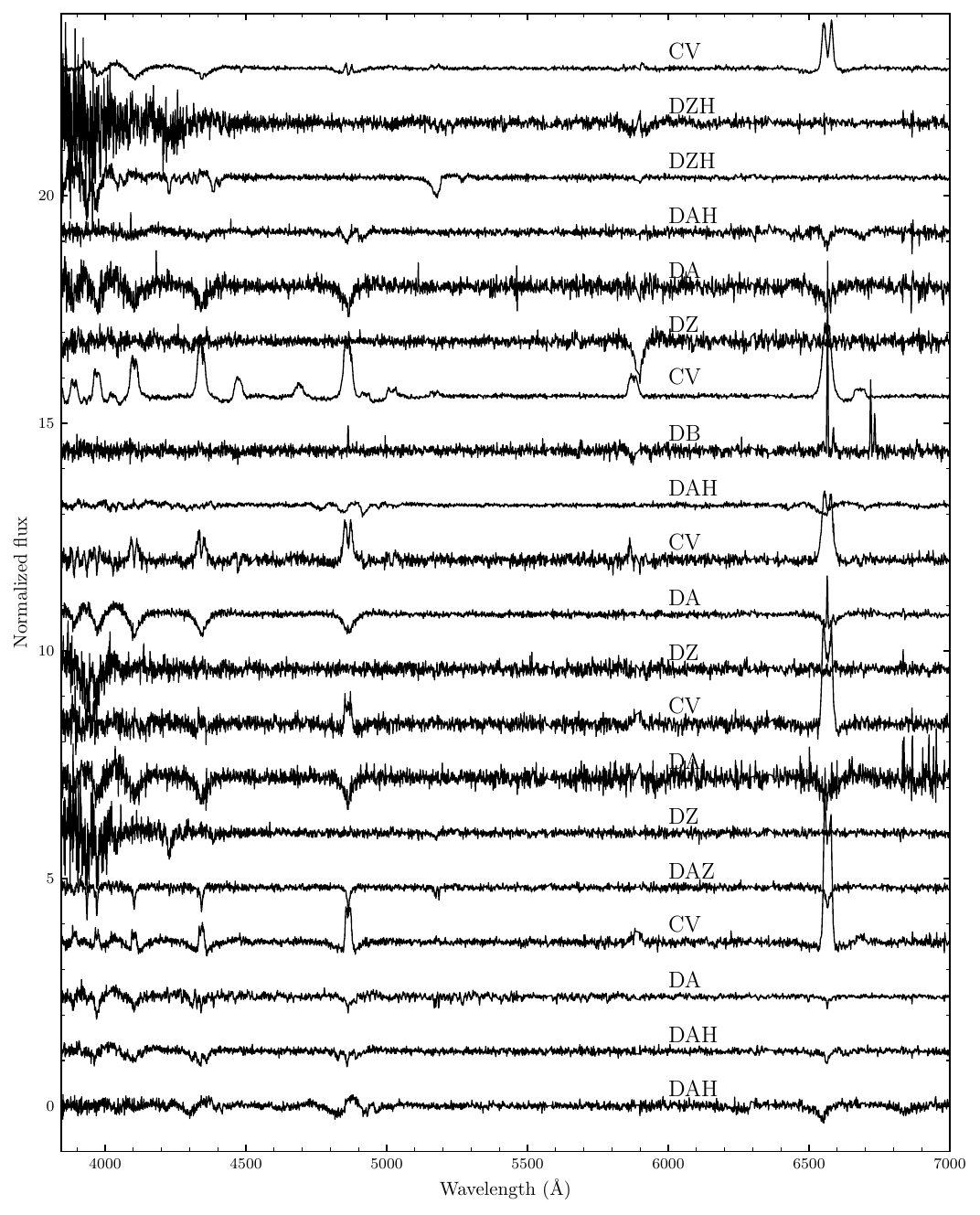}
    \caption{Random selection of spectra tagged as outliers in Section \ref{sec:outliers}. Flux is continuum-normalized and arbitrarily shifted for clarity. The spectral type from GF21 is indicated as a reference.}
    \label{fig:outliers}
\end{figure*}


\bsp	
\label{lastpage}
\end{document}